\documentclass[aps,prl,twocolumn,showpacs]{revtex4}
\usepackage{graphicx}
\usepackage{amssymb}

\begin{document}

\title{Sublattice addressing and spin-dependent motion of atoms in a double-well lattice}

\author{P. J. Lee, M. Anderlini\footnote{Present address: INFN, Sezione Di Firenze, via Sansone 1, I-50019
Sesto Fiorentino (FI), Italy
}, B. L. Brown, J. Sebby-Strabley\footnote{Present address: Honeywell Aerospace, 12001 State Highway 55, Plymouth, MN, 55441}, W. D. Phillips,  and J. V. Porto\footnote{Electronic address: trey@nist.gov
} }

\affiliation{Joint Quantum Institute, National Institute of Standards and Technology
and University of Maryland, Gaithersburg MD, 20899 USA}

\date{\today}

\begin{abstract}
We load atoms into every site of an optical lattice and selectively spin flip atoms in a sublattice consisting of every other site.  These selected atoms are separated from their unselected neighbors by less than an optical wavelength.  We also show spin-dependent transport, where atomic wave packets are coherently separated into adjacent sites according to their internal state. These tools should be useful for quantum information processing and quantum simulation of lattice models with neutral atoms.
\end{abstract}


\maketitle
Ultracold neutral atoms in optical lattices offer new possibilities
for the study of condensed matter systems and quantum information
processing \cite{Bloch2005,Bloch06,Brennen1999,Calarco2004,Jaksch1999}.
One of the outstanding issues regarding the use of this system for
quantum computing is the ability to address individual qubits. In principle, this can be accomplished using focussed laser beams \cite{Zhang2006}, and progress has been made toward solving this problem \cite{bergamini04,Darquie2005,dumke02,Schrader2001,Miroshnychenko,Weiss2004}.  However, control of internal spins with spatial resolution on the scale of an optical wavelength or less remains a challenge.

Here we describe experiments with an optical lattice of double-well potentials
in which we address and manipulate atoms in one site of every double well, independently of neighboring sites at a distance of $\lambda/2$, where $\lambda=803$\,nm is the wavelength of the light creating the optical lattice. 
A spin-dependent (i.e.~internal-state dependent)
light shift provides a local effective magnetic field gradient sufficiently
large to spectroscopically resolve spin-flip transitions \cite{Stokes1991} at neighboring sites.  
While we simultaneously address atoms in every other site (instead of only in an individual site), the basic principle demonstrated here underlies proposals to use state- and position-dependent light shifts to address individual atoms \cite{Zhang2006}.
Furthermore, by manipulating the optical lattice potential we induce coherent spin-dependent
transport, resulting in spatial separation of different spin states into neighboring sites.
The control techniques presented here using the double-well lattice provide the tools necessary for exploring two-qubit entangling gate schemes based on coherent collisions \cite{Jaksch1999} or exchange interactions \cite{Deutsch2007} between two atoms.  
The ability to independently control atoms in a sublattice also expands the accessible parameters for quantum simulations, for example, by creating a patterned loading of the lattice \cite{Zhang2007}, or simulating Hamiltonians stroboscopically \cite{Jane2003}.

\begin{figure}[t]
\centering

\begin{centering}
\includegraphics[width=3.0in,keepaspectratio]{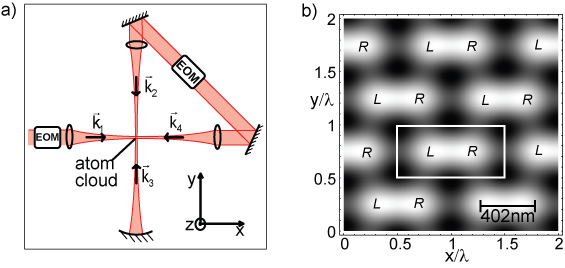} 
\end{centering}

\caption{a) Schematic of the 2D horizontal lattice setup.  b) Example of a double-well optical lattice potential in the horizontal plane.  Each copy of the unit cell (indicated by the rectangular box) is an isolated system, and atoms in the $L$ sublattice can be addressed by rf independently from atoms in the $R$ sublattice (and vice versa).}

\label{schematic} 
\end{figure}

\begin{figure}[t]

\begin{centering}
 \includegraphics[width=3.2in,keepaspectratio]{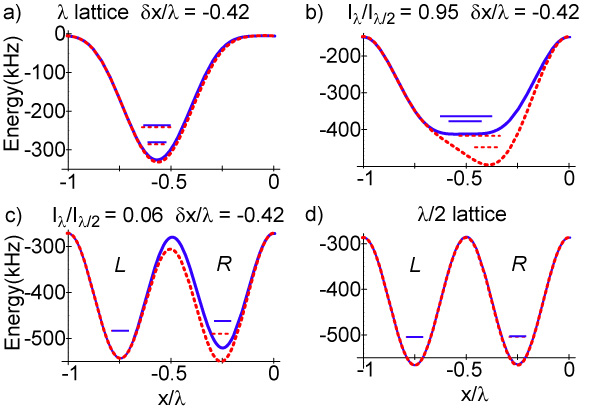}
\end{centering}
\caption{Calculated double-well potentials along $x$ for different configurations (a)--(d) of the laser light. Dashed lines represent the scalar potential
$U_{{\rm s}}$ seen by atoms in the $|m_{F}=0\rangle$ state, solid
lines show the state-dependent potential $U_{{\rm s}}+U_{{\rm v}}$
seen by atoms in the $|m_{F}=-1\rangle$ state. (The large constant
Zeeman shift has been subtracted.) The energy levels for the first
two vibrational eigenstates of each spin are also shown as horizontal
lines, with the position and the length of the lines representing
the center-of-mass position and the rms full width of the corresponding
wavefunctions.  ($\delta x/\lambda=-0.5$ corresponds to the $\lambda$-lattice minima exactly half way between the $\lambda/2$-lattice minima.)}

\label{potentials} 
\end{figure}

In our experiment, we create a 3D lattice having a unit cell that can be dynamically transformed between single well and double well configurations \cite{sebby06,anderlini06}.  The 2D lattice in the horizontal $xy$ plane is formed by intersecting two orthogonal pairs of counter-propagating laser beams derived from a single, folded,
retro-reflected beam (Fig.~\ref{schematic}), while an independent 1D optical lattice provides confinement in the vertical direction $z$.  For the 2D lattice, horizontally polarized light forms a ``$\lambda/2$-lattice'' with a periodicity of $\lambda/2$ along both $x$ and $y$ (Fig.~\ref{potentials}-d). Vertically polarized light forms a ``$\lambda$-lattice'' with periodicity
$\lambda$ along $x$ (Fig.~\ref{potentials}-a).
Their relative intensity $I_{\lambda}/I_{\lambda/2}$ and relative
position $\delta x$ along $x$ are independently controlled
by manipulating the polarization and phases of the beams using electro-optic modulators (EOMs).
For the experiments presented here, we focus only on dynamics within each double well, and configure the lattice so that tunneling between separate double wells is negligible on the timescale of the experiment.
Configurations resulting in local elliptical
polarization (i.e.~a complex polarization vector) produce a spin-dependent
potential, originating from the vector light shift $U_{{\rm v}}=i\alpha_{{\rm v}}(\mathbf{E}^{*}\times\mathbf{E})\cdot\mathbf{F}/4$,
where Re{[}${\mathbf{E}}e^{-i\omega t}$] is the laser electric field,
$\alpha_{{\rm v}}$ is the vector part of the atomic polarizability,
and $\hbar\mathbf{F}$ is the total atomic angular momentum \cite{Deutsch1998}.
The vector light shift $U_{{\rm v}}$ can be treated as arising from
an effective magnetic field $\mathbf{B}_{{\rm eff}}(\mathbf{r})\propto i\alpha_{{\rm v}}(\mathbf{E}^{*}(\mathbf{r})\times\mathbf{E}(\mathbf{r}))$.
In the presence of a uniform bias field $\mathbf{B}_{{\rm 0}}$,
the total field is $\mathbf{B}_{\rm tot}(\mathbf{r})=\mathbf{B}_0+\mathbf{B}_{\rm eff}(\mathbf{r})$.
Atoms in different Zeeman levels experience different vector
light shifts, in addition to the scalar optical potential $U_{{\rm s}}=-\alpha_{{\rm s}}|\mathbf{E}|^{2}/4$,
which depends on the intensity but not on the polarization. 

For these experiments, we use the ground state $5S_{1/2}$, $F=1$
manifold of $^{87}$Rb. A bias field $\mathbf{B}_{{\rm 0}}$ points in
the direction of $\hat{x}-\hat{y}$ and creates a sufficiently large
quadratic Zeeman shift such that the radio frequency (rf) can be tuned to selectively
couple only $|m_{F}=-1\rangle$ and $|m_{F}=0\rangle$, which represent
our qubit states.  
(In principle the qubit states can be any two levels with a magnetic-field dependent energy splitting.) 
Fig.~\ref{potentials} shows several examples
of the potential for $^{87}$Rb atoms in these two states,
experimentally attainable with 100\,mW of light at $\lambda=803$\,nm
focused to a $1/e^{2}$ radius of 170\,$\mu$m. For this lattice, state-dependence (when it exists) is predominantly
in the right well (see e.g.~Fig.~\ref{potentials}-c). For purely vertical or horizontal polarization (Figs.~\ref{potentials}-a and \ref{potentials}-d), $U_{{\rm v}}$ is zero. In contrast, simultaneous
presence of both polarizations can
result in a $U_{{\rm v}}$ as large as 100\,kHz for the $|-1\rangle$
state ($|0\rangle$ is not sensitive to magnetic fields) %
\footnote{The states labeled $|0\rangle$ and $|-1\rangle$ are energy eigenstates
of the atoms in their local optical and magnetic fields. %
}. For a configuration such as in Fig.~\ref{potentials}-c,
the transition frequency from $|-1\rangle \rightarrow |0\rangle$
is different for atoms in the right ($R$) and left
($L$) sites. We use this difference to selectively
rf-address atoms in either the $R$ or $L$ sublattice.

We begin the experiment by creating a magnetically trapped Bose-Einstein
condensate of 2.0(5)$\times10^{5}$ $^{87}$Rb atoms in the
$|-1\rangle$ state \cite{Peil2003} and loading it into the $\lambda/2$-lattice
(Fig.~\ref{potentials}-d). The double-well and vertical lattices are turned on together in 500\,$\mu$s with an exponential ramp \cite{anderlini06} (having a 100\,$\mu$s
time constant) in order to adiabatically
populate only the ground band of the optical potential.  (We did not load more slowly to obtain a number-squeezed Mott insulator state \cite{Greiner2002} since the single particle effects explored in these experiments are largely insensitive to atom-atom interactions.) The final depth of the double-well and the vertical lattices are about 80$E_{{\rm R}}$ and 52$E_{{\rm R}}$
respectively, where $E_{{\rm R}}=h^{2}/(2m\lambda^{2})$=\,3.5\,kHz is the recoil
energy and $m$ is the rubidium mass. After loading, the
magnetic trapping fields are turned off, except for a
bias field $|\mathbf{B}_{{\rm 0}}|$ of $\approx$4.8\,mT (giving a 34\,MHz linear- and a 300\,kHz
quadratic-Zeeman shift).  We wait 40\,ms for the field to stabilize.

\begin{figure}[t!]
\centering
\begin{centering}
\includegraphics[width=2.8in,keepaspectratio]{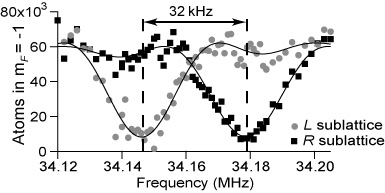} 
\end{centering}
\caption{Measured population in $|-1\rangle$ in the $L$ (circles) and $R$ (squares) sublattice after the rf pulse.  The lines are fits to the expected lineshape. 
The resonant frequency in the spin-dependent $R$ sublattice (34.178\,MHz) is shifted by the vector light shift from the frequency in the spin-independent $L$ sublattice (34.146\,MHz). }

\label{spectroscopy} 
\end{figure}

To address atoms in either the $L$ or $R$ sublattice, we change the lattice configuration to the spin-dependent one
shown in Fig.~\ref{potentials}-c, and apply a 30\,$\mu$s rf pulse
coupling $|-1\rangle$ to $|0\rangle$. The number of atoms remaining in $|-1\rangle$
in each sublattice is then measured using the following procedure:
the magnetic trap is turned back on, the vertical lattice is turned
off, and the double-well lattice potential is reduced to 30$E_{{\rm R}}$.
This reduces the vertical confinement for atoms in $|0\rangle$ such
that they fall out of the trap (due to gravity), while keeping the atoms in $|-1\rangle$
in their original $L$ and $R$ sites. After 10\,ms, we return the
double well lattice to 52$E_{{\rm R}}$ depth. Finally, we measure
the remaining  population in each sublattice by adiabatically adjusting
the lattice so that atoms in the $R$ sublattice transform to highly excited
vibrational states while atoms in the $L$ sublattice remain in the ground
vibrational state, as described in ref \cite{Sebby-Strabley2007}.
Fig.~\ref{spectroscopy} shows a measurement of the residual population
in state $|-1\rangle$ in the two sublattices as a function of the rf frequency. The two resonances differ by $32$\,kHz,
which is well-resolved by the 30\,$\mu$s rf pulse and is in good
agreement with a calculation based on the effective magnetic field.
We have also demonstrated sublattice-resolved hyperfine transitions driven
by microwaves, and in principle could use optical Raman fields as
well.

The spin-dependent double-well also allows for state-dependent motion
of the atomic wave packets, entangling spatial and spin degrees of freedom.
This is similar to the controlled spin-dependent motion demonstrated
in, e.g., refs \cite{mandelNature03,Jessen2000}, but here the double-well
lattice breaks left-right symmetry within each pair of sites, confining
motion to the unit cell containing those two sites. In our transport
experiment, each atom initially in the ground state of the $\lambda$-lattice
is sorted into the $L$ or $R$ well of the $\lambda/2$-lattice
according to its spin. To do this we simultaneously control
the total intensity, the ratio $I_{\lambda}/I_{\lambda/2}$, and $\delta x$
in a two-step process. In the first step, we start with a $\lambda$-lattice
configuration (similar to Fig.~\ref{potentials}-a) having a depth
of 100$E_{{\rm R}}$ and $\delta x=-0.62\lambda$. This is then transformed
to a potential similar to Fig.~\ref{potentials}-b in 300\,$\mu$s,
ending with a $\sim100E_{\rm{R}}$ potential depth. The final intensity ratio is
$I_{\lambda}/I_{\lambda/2}=0.95$ and the minimum of the $\lambda$-lattice
is located in the right well (for $\delta x/\lambda>-0.5$) . In the second
step, beginning at Fig.~\ref{potentials}-b, the barrier is raised
(Fig.~\ref{potentials}-c) at fixed $\delta x/\lambda$ to separate the two spin states into the
$L$ and $R$ sites, by fully transforming into the $\lambda/2$-lattice
(Fig.~\ref{potentials}-d) in 100\,$\mu$s.
The population in the $L$ and $R$ sublattices are then measured as above
\cite{Sebby-Strabley2007}. 

\begin{figure}[t!]
\begin{centering} 
\includegraphics[width=3.0in,keepaspectratio]{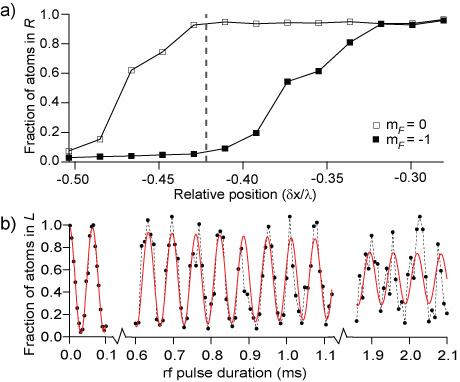}
\end{centering}

\caption{a) Fraction of atoms in the $R$ sublattice at the end of the spin-dependent transport sequence (see text) for atoms
in the $|-1\rangle$ state (filled) and $|0\rangle$
state (open) as a function of the relative position
$\delta x/\lambda$ between the $\lambda$-lattice and the $\lambda/2$-lattice.
The dashed vertical line indicates the best selective transfer:
$\delta x=-0.42\lambda$, where atoms in $|0\rangle$ are sorted
into $R$ sites and atoms in $|-1\rangle$ are sorted into $L$
sites, each with 92(2)\% probability. b) Using spin-dependent transport to measure spin populations after applying an rf pulse to atoms in the $\lambda$-lattice.  Atoms in $|-1\rangle$ are in the $L$ sublattice after transport.  During the rf pulse, atoms oscillate between $|-1\rangle$ and $|0\rangle$. The red curve is a fit to a damped sine.}

\label{SelectiveSwap} 
\end{figure}

To determine the best final value for $\delta x$, the sequence is
performed on atoms in the $|-1\rangle$ and $|0\rangle$
states in separate trials. In Fig.~\ref{SelectiveSwap}, the fraction of atoms in the $R$ sublattice after the transfer sequence is
plotted as a function of $\delta x/\lambda$.  $U_{{\rm v}}$ creates a difference in the
potentials for the two Zeeman levels, which for a narrow range of $\delta x/\lambda$, transfers atoms in $|0\rangle$ to $R$ with almost unit efficiency while
atoms in $|-1\rangle$ remain in $L$. At $\delta x=-0.42\lambda$,
the atoms are localized to the site corresponding to their spin with
92\% probability, which includes imperfections in preparing the spin states, state-dependent transport, and the population measurement. (Systematic measurement error limits our measured fidelity to 96\%.)  As with any process requiring adiabatic motion, there is a trade-off between fidelity and speed, though in principle, 100\% fidelity is possible if other decoherence mechanisms do not limit the time allowed for the operation.  The range of parameters resulting in spin-dependent separation, and the fidelity and speed of the process can be improved by increasing the lattice depth (thereby reducing the time scale required for adiabaticity).  Optimal quantum control techniques \cite{Calarco2004} are also expected to reduce the transfer time while maintaining fidelity \cite{DeChiara}.

Spin-dependent transport can act as a microscopic Stern-Gerlach filter,
which we use here to study spin decoherence due to the magnetic
field environment.  Spin dephasing in our experiments is dominated by (1) field fluctuations between data points (separated by $\approx$1 minute), (2) field fluctuations during a single shot ($\lesssim$ a few ms), and (3) spatial inhomogeneity in the field (across the $\sim$35\,$\mu$m atom cloud).  First, we observe Rabi oscillations by applying an rf pulse resonant with the $|-1\rangle\rightarrow|0\rangle$ transition for atoms in the spin-independent $\lambda$-lattice and then use spin-dependent transport to detect the final spin populations (see Fig.~\ref{SelectiveSwap}-b).
All three noise sources act as an effective detuning inhomogeneity, to which the averaged Rabi cycling is only sensitive quadratically \cite{Allen-Eberly}.  The data shows a Rabi frequency of 15.8\,kHz and a $1/e$ decay time of 3\,ms (25 Rabi flops).  

We also performed a Ramsey spin-resonance experiment \cite{Allen-Eberly} by applying a resonant $\pi/2$ rf pulse that places the atoms in a superposition of spin states, followed by a second $\pi/2$ pulse with the same frequency and a relative phase $\theta$ to read out the coherence after a given delay.  
The Ramsey interferogram is obtained from successive measurements with different $\theta$ for each delay time. 
The contrast of the interferogram decays due to mechanisms with two different timescales.   The first (100\,$\mu$s)  is related to a random phase from shot to shot that is the same for all atoms in the sample.  It results from noise sources (1) and (2) and is indicated by a random population transfer on any given shot that fluctuates between $\sim$0 and $\sim$100\%.  The second (500$\mu$s) is related to inhomogeneity across the sample (3), and is indicated by population transfers tending toward 50\% with little fluctuation.  We also perform a spin echo experiment by inserting a $\pi$ pulse symmetrically between the two $\pi/2$ pulses, which effectively reverses the dephasing caused by (1) and (3).  Here, the first timescale (due to the remaining shot-to-shot fluctuations caused by (2)) increases to 400\,$\mu$s, and the second increases to at least 1.2\,ms.  Noise sources (1) and (2) can be improved by active stabilization of the magnetic field.

\begin{figure}[t!]
 \includegraphics[width=2.9in,keepaspectratio]{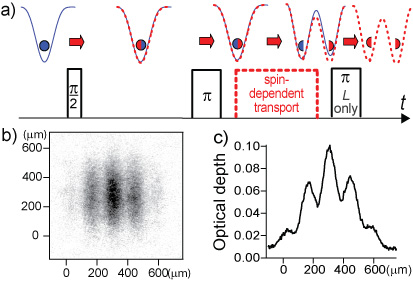}
\caption{\label{Coherent Swap Sequence} a) Experimental sequence for demonstrating
coherence in spin-dependent dynamics. Following the first $\pi$ pulse,
wavefunctions are spin-dependently split into $L$ and $R$ sites.
A $\pi$ pulse is applied to atoms in $L$ sites at
the end of the transport so that all atoms end in the
same final spin state. b) The double-slit diffraction observed after
the lattice is snapped off and after time-of-flight, indicating
coherence after spin-dependent transport. c) Integrated density profile
of the double-slit diffraction pattern in (b).}
\end{figure}

To investigate whether coherence is preserved during the spin-dependent
transport, we use an atom interferometer similar to that of ref \cite{Sebby-Strabley2007}.
In this case, however, the spin-dependent transport acts as an analog
to an optical {\em polarizing} beam splitter,
separating atoms initially in the ground state of the $\lambda$-lattice
into the two output modes $|L\rangle$ and $|R\rangle$ depending
on their $m_{F}$ state. The experimental
sequence is shown in Fig.~\ref{Coherent Swap Sequence}-a. Atoms loaded
in the $\lambda$-lattice are first placed in a superposition of spin
states: $\left(|-1\rangle+|0\rangle\right)/\sqrt{2}$, using a rf
$\pi/2$ pulse. Since the spin-dependent transport time (370\,$\mu$s)
is comparable to the inhomogeneous spin coherence time, we perform
the transport inside a spin echo sequence. The echo-inducing $\pi$
pulse is applied 400\,$\mu$s after the $\pi/2$ pulse. The atoms
are then separated through spin-dependent transport, which
entangles each atom's spatial wavefunction with its spin, creating
the single particle state $\left[\left(|-1\rangle\otimes|L\rangle\right)+e^{i\phi}\left(|0\rangle\otimes|R\rangle\right)\right]/\sqrt{2}$.
At this point, atoms in the two paths of the interferometer are in
orthogonal spin states. Therefore,
we apply a $\pi$ pulse resonant only in the $L$ sublattice, forming the state $|0\rangle\otimes\left(|L\rangle+e^{i\phi}|R\rangle\right)/\sqrt{2}$.
Fig.~\ref{Coherent Swap Sequence}-b is an example of the double-slit
diffraction pattern observed when the atoms are released. (Due to
interactions, the visibility of the interference fringes undergoes
collapse and revival, as in ref \cite{Sebby-Strabley2007}; Fig.~\ref{Coherent Swap Sequence}-b
is taken at a revival, thereby removing the effect of interactions on the diffraction pattern.) The clear appearence of a double-slit diffraction
pattern is an indication that some coherence is preserved during spin-dependent
transport. The visibility of the fringe pattern is less than the best
we have seen for spin-independent interferometry \cite{Sebby-Strabley2007}, likely due to decoherence during the spin echo sequence and off-resonant coupling to atoms in the $R$ sublattice in the last $\pi$-pulse.

If the spin-dependent transport experiment is performed in reverse, i.e.~atoms initially in two separate wells are brought together into a single well through spin-dependent transport, the atoms could become entangled through contact interaction.  In this scenario, the ability to selectively address atoms in two neighboring sites would allow us to gain access to the full two-qubit Hilbert space (assuming we start with a Mott insulator state).  If the principles presented here are applied using focused light beams \cite{Zhang2006}, these universal operations within our two-qubit toy system could be scalable to large numbers of atoms.  However, much work is still needed to explore ideas to generate and control two-atom interactions.  The flexibility of the double-well optical lattice provides a testbed for some of these ideas before they are implemented in a scalable architecture.

This work was partially supported by DTO, ONR, and NASA. The authors
warmly thank Jens Kruse, Ian Spielman, and Steve Rolston for helpful
contributions. PJL, BLB, and JS-S acknowledge support from the NRC.

\end{document}